\documentclass[twocolumn]{revtex4}

\newcommand {\half}{\frac{1}{2}}

\newcommand {\eighth}{\frac{1}{8}}

\newcommand {\ff}{f}
\newcommand {\pp}{p}

\begin{document}

\title
{Beyond the Child-Langmuir Limit}

\author{R.E. Caflisch$^{1,2}$, and M.S. Rosin$^{1}$}

\affiliation{$^1$Department of Mathematics, UCLA, Los Angeles, CA 90095 \\
$^2$Institute for Pure and Applied Mathematics, UCLA, Los Angeles,
CA 90095 }

\begin{abstract}

This paper describes a new solution formulation for  fully nonlinear and unsteady planar flow of an electron beam in a diode.  Using characteristic variables - i.e., variables that follow  particle paths - the solution is expressed through an exact analytic, but implicit, formula for any choice of  
incoming velocity $v_0$, electric field $E_0$ and current $J_0$. For steady solutions, this approach clarifies the origin of the maximal  current $J_{max}$, derived by Child and Langmuir for  $v_0=0$ and by Jaffe for $v_0>0$. The implicit formulation is used   to find (1) unsteady solutions   having constant incoming flux $J_0>J_{max}$, which leads formation of a virtual cathode, and (2) time-periodic solutions whose  average flux exceeds the adiabatic average of $J_{max}$.

\end{abstract}

\maketitle
Space charge limiting (SCL) current is a fundamental constraint on the flow of an electron beam in a diode. For  fixed potential difference $|\phi_1|$ 
and incoming velocity $v_0$, the maximal sustainable current $J_{max}$ was derived by Child \cite{child} and Langmuir \cite{langmuir} for $v_0=0$ and by Jaff\'{e} \cite{jaffe} for $v_0>0$. The physics origin of the SCL effect is clear:  electromotive force from electrons in the diode limits the current in the diode. If  incoming current is  maintained above this maximum, then the electron density builds up inside the diode and a virtual cathode develops. 

The mathematical  derivation of the maximal current in  \cite{child,langmuir,jaffe} is based on equations for the steady, one-dimensional electron flow in a diode. The authors derive  a formula relating the current and the potential jump, but 
the analysis for $v_0>0$ in \cite{jaffe} is quite complicated. On the other hand, formation of a virtual cathode is known to be due to cusp formation in the electron trajectories \cite{CoutsiasJPP1984,CoutsiasJPP1988,CoutsiasSullivan}. The basic physics and technological applications of SCL flows and virtual cathodes are well reviewed in \cite{BirdsallBridges,miller}. Extensions to more general physics and geometries have been carried out, mostly using perturbation methods, e.g., \cite{Luginsland2002} for multi-dimensional geometries.

This paper presents a new formulation for the complete solution of the one-dimensional diode equations. The solution is based on characteristics (i.e., particle paths) so that it is an extension of \cite{CoutsiasJPP1984,CoutsiasJPP1988,CoutsiasSullivan}, but it applies to both  steady and unsteady flow and to  the fully nonlinear equations with no approximations. 

The result is an implicit solution, analogous to the implicit solution for the inviscid Burgers equation (e.g., see \cite{lax}), since velocity $v$, density $\rho
$, electric field $\psi$, and spatial position $x$ are given in terms of characteristic variables $s$ and $\tau$. From this implicit formulation, it is staightforward to derive the maximal current that was first found by \cite{child,langmuir,jaffe} and the mathematical meaning of the derivation is clearer in this formulation: There is a single derivation for $v_0=0$ and $v_0>0$; the steps in the derivation are integrals along particle paths; and the maximal value of the flux is interpreted as a minimal value for the potential  as a function of the incoming electric field (for fixed  flux). 
 In addition, the implicit solution formulation enables construction of unsteady solutions that exhibit important properties, including 
singularity formation corresponding to cusp formation in the characteristics and formation of a virtual cathode, and time-periodic solutions whose  average flux exceeds the adiabatic average of $J_{max}$.   

The one-dimensional continuum equations for the flux of electrons in a diode are 
\begin{eqnarray}
\partial_t\rho &+& \partial_x (\rho v) = 0  \label{mass}\\
\partial_t v &+& v\partial_x v = - \partial_x \phi \label{momentum} \\
\partial_{x}^2 \phi &=& - \rho \label{poisson}
\end{eqnarray}
in which $x,t,v,\phi,\rho$ are the scaled position, time, velocity, potential and density given by
\begin{eqnarray}
(x,t,v,\phi,\rho) = (x'/L, t'/T, v/(L/T), \phi'/\Phi, \rho'/R) \nonumber \\
\Phi = (m_e/q_e) (L/T)^2 \quad \quad R=\varepsilon_0 \Phi/(q_e L^2) . \nonumber 
\end{eqnarray}
The primed variables are unscaled, $L,T$ are characteristic length and time values,  $ m_e,q_e$ are electron mass and charge and $\varepsilon_0$ is vacuum permittivity. 
The boundary conditions at the cathode  $x_0=0$  and anode  $x_1=d$  are
\begin{eqnarray}
\left. \begin{array}{lll}
\phi = 0 \\ v=v_0 \\ \rho = \rho_0
\end{array}  \right\} \ \hbox{on} \ x = 0  \label{leftbdry}
\\
\phi = \phi_1 \ \ \hbox{on} \ x = d . \nonumber 
\end{eqnarray}
so that $-\phi_1$ is the potential difference across the channel (the negative sign is used since $\phi_1<0$).

Consider characteristic (particle path) variables in which $x(s,\tau)$ is the position at time $t=s+\tau$ for a particle that entered the domain at time $\tau$ and is  moving at speed $v$. The defining equations for $s$ and $\tau$ are
\begin{eqnarray}
\partial_s x &=& v \label{charx}\\
x(0,\tau)&=&0 \label{bdryx} \\
t&=&s+\tau .\nonumber 
\end{eqnarray}
Derivatives in $(x,t)$ and in $(s,\tau)$ are related by
\begin{eqnarray}
\partial_s&=& \partial_t + v \partial_x \label{sderiv} \\
\partial_\tau &=& \partial_t +(\partial_\tau x) \partial_x . \nonumber 
\end{eqnarray}

Denote (scaled) electric field by $\psi=-\partial_x \phi$. Since $\partial_x \psi=\rho$, then $\psi(x,t)$ is the total mass between $0$ and $x$, plus some boundary terms, which implies
\begin{equation}
\partial_t  \psi + v\partial_x \psi=\ff'''' (t) \label{psi1}
\end{equation}
for some function $\ff''''$ (the four derivatives are for notational convenience below).  Combine eq.~(\ref{psi1}) with eq.~(\ref{momentum}) and eq.~(\ref{charx}), using eq.~(\ref{sderiv}), to get the following system
\begin{eqnarray}
\partial_s \psi&=& \ff ''''(s+\tau) \nonumber \\ 
\partial_s v &=& \psi  \label{charform} \\ 
\partial_s x&=& v .  \nonumber  
\end{eqnarray}
The general solution for this system, using eq.~(\ref{bdryx}),  is 
\begin{eqnarray}
\psi(s,\tau)&=& \theta(\tau)+\ff '''(s+\tau) \nonumber \\ 
 v(s,\tau) &=& w(\tau)+\theta(\tau)s+\ff ''(s+\tau) \label{charsltn1} \\ 
 x(s,\tau)&=& w(\tau)s+\half \theta(\tau)s^2+\ff '(s+\tau)-f'(\tau). \nonumber  
\end{eqnarray}
The  system~(\ref{charsltn1}) provides a  new  general method for solving the unsteady  diode eqs.~(\ref{mass})-(\ref{poisson}). 

In eq.~(\ref{charsltn1}) $\ff$, $\theta$ and $w$ are related to  boundary data 
by
\begin{eqnarray}
\ff ''''&=& \partial_\tau \psi_0 + J_0 \nonumber \\ 
 \theta &=& \psi_0 - \ff ''' \nonumber \\ 
 w&=& v_0-\ff ' . \nonumber  
\end{eqnarray}
in which $J_0=\rho_0 v_0$ and $\psi_0=\psi(x=0)$ are  incoming flux and electric field.
Specification of boundary data on $x=d$ requires identification of the crossing time $s=T(\tau)$ at which characteristics (particle paths) hit  $x=d$; i.e. 
\begin{eqnarray}
 x (T(\tau),\tau) =d.\nonumber  %
\end{eqnarray}

The density, flux and potential satisfy (using $\partial_x=(v-\partial_\tau x )^{-1} (\partial_s - \partial_\tau )$ and $\partial_\tau x=0$ at $x=0$)
\begin{eqnarray}
\rho&=&  (v-\partial_\tau x )^{-1} (\partial_s - \partial_\tau ) \psi  \label{rhoeqtn}\\ 
J&=&  v (v-\partial_\tau x )^{-1} (\partial_s - \partial_\tau ) \psi  \nonumber \\ 
J_0 &=&(\partial_s - \partial_\tau ) \psi(0,\tau)  \nonumber \\ 
(\partial_s - \partial_\tau ) \phi &=&-(v-\partial_\tau x) \psi. \label{phieqtn} 
\end{eqnarray}
Eq.~(\ref{phieqtn})  can be integrated  (using $\phi(0,\tau)=0$) to get
\begin{equation}
 \phi (s,\tau) = - \int_0^s (v-\partial_\tau x) \psi (s',\tau+s-s') ds'. \label{phiIntegral}
\end{equation}

Next consider steady solutions, which  cannot depend on $\tau$, so that $\ff ''''=J_0$ is a constant, and the resulting solutions of system~(\ref{charform}) are
\begin{eqnarray}
\psi(s)&=& \psi_0+J_0 s  \nonumber \\ 
 v(s) &=& v_0+\psi_0 s+J_0 s^2/2   \nonumber \\ 
 x(s)&=& v_0 s+\psi_0 s^2/2+J_0 s^3/6   \label{steadyeqtn} \\
\phi(s)&=& \half(v_0^2 - v(s)^2).  \nonumber  
\end{eqnarray}
In particular, the value $\phi_1$ of the potential at $x=d$ is
\begin{equation}
\phi_1 =  \half (v_0^2 - v(T)^2)=\half v_0^2 - \half ( v_0 + \psi_0 T + \half J_0 T^2) ^2. \label{phi1eqtn}
\end{equation}

This is equivalent to the solutions derived by Child \cite{child,langmuir,jaffe}. They showed that there is maximal value $J_{max}$ of the current for given values of the potential difference $-\phi_1$, or equivalently that there is a minimal value $-\phi_{min}$ of the  potential difference $-\phi_1$ for given values of the incoming current $J_0$.

The characteristic solution eq.~(\ref{steadyeqtn}) shows that  there is a solution of the diode equations  for any choice of the mathematically natural boundary data $v_0$, $J_0$ and $\psi_0$. 
The reason for  the minimal potential jump $-\phi_{min}$ (or equivalently the maximal current $J_{max}$) is just that the function $-\phi_1$ has a minimum value  as $\psi_0$ is varied, for fixed values of velocity $v_0$ and current $J_0$.

In order to find this minimum value, first calculate $\partial_{\psi_0} T$ by differentiating the equation $ v_0 T+ \psi_0 T^2/2 +  J_0 T^3/6 =d$ and eq.~(\ref{phi1eqtn}) with respect to $\psi_0$ to get
\begin{eqnarray}
\partial_{\psi_0} T &=& -\half v(T)^{-1} T^2  \nonumber \\ %
\partial_{\psi_0} \phi_1 &=& -T(v_0 + \half \psi_0 T) .  \nonumber  %
\end{eqnarray}
The minimal value of $-\phi_1$ occurs when $\partial_{\psi_0} \phi_1=0$ which implies 
\begin{eqnarray}
\psi_0 &=&  -2 v_0 T^{-1}  \nonumber \\ 
 d&=&JT^3/6 . \nonumber 
\end{eqnarray} 
At this value of $T$, the potential difference  $-\phi_1 = -\phi_{min}$ and current $J_0=J_{max}$  are 
\begin{eqnarray}
\phi_{min} &=& \half  v_0  (36 d^2 J_0)^{1/3} - \eighth (36 d^2 J_0)^{2/3}    \nonumber \\ 
J_{max} &=& \frac{2}{9} d^{-2}\left( v_0 + \sqrt{v_0^2-2 \phi_1}\right)^3. \label{Jmax}
\end{eqnarray}
which is equivalent to the results of  \cite{child,langmuir,jaffe}.

The allowable range of the parameter $\psi_0$ for steady solutions is determined from the following two requirements: The potential difference is negative (i.e., $\phi_1<0$) so that electrons are driven to the anodes from the cathode; and
the velocity is always positive (i.e., $v(s)>0$ for $0<s<T$) since otherwise the particle paths are crossing and the model breaks down. These imply $J_0>0$, $v_0>0$ and $-J_0 T/2 < \psi_0$. Stability analysis is easily performed using the implicit solution and shows that the steady solution is stable for $\psi_0>-2v_0/T$, unstable for $-2v_0/T>\psi_0>-\sqrt{2J_0 v_0}$ and nonexistent for $-\sqrt{2J_0 v_0}>\psi_0$. These results are consistent with, but more easily stated than those of \cite{jaffe}.

Next we construct several unsteady solutions  that exhibit interesting behavior. 
As a first example,  consider unsteady solution  having constant incoming velocity $v_0$ and flux $J_0$. The implicit solution then has the form
\begin{eqnarray}
\psi(s,\tau)&=& \theta_0+J_0 s +\ff '''(s+\tau)   \nonumber \\ 
 v(s,\tau) &=& v_0+\theta_0 s + J_0 \half s^2+\ff ''(s+\tau)-\ff ''(\tau) \nonumber \\ 
 x(s,\tau)&=& v_0s+\half \theta_0 s^2+ J_0 \frac{1}{6} s^3+\ff '(s+\tau)-\ff '(\tau)-\ff ''(\tau)s .  \nonumber  
\end{eqnarray}
The corresponding potential $\phi$ is found explicitly  as 
\begin{eqnarray}
 -\phi (s,\tau)
 &=&(\theta_0+\ff '''(s+\tau)) x(s,\tau) + \pp_3(s)-d\theta_0 \nonumber \\
 && + 2J_0(-\ff (\tau)+\ff (\tau+s)-s\ff '(\tau)-\frac{1}{2} s^2 \ff ''(\tau) ) \nonumber 
\end{eqnarray}
in which $\pp_3$ is defined by eq.~(\ref{p3}). At  $s=T(\tau)$, the equations for $\phi_1(\tau)=\phi (T,\tau)$ and $T(\tau)$  become
\begin{eqnarray}
 -\phi_1(\tau)
 &=& d \ff '''_+  +   \pp _3(T) + 2J_0(-\ff +\ff _+ -T \ff '-\frac{1}{2} T^2 \ff '' ) \nonumber \\
d &=&\pp _1(T) - \ff ''T + \ff '_+ - \ff ' \label{eqtnT1}
\end{eqnarray}
in which
\begin{eqnarray}
 \pp _1 (T)&=&v_0 T + \theta_0 \frac{1}{2} T^2 +  J_0 \frac{1}{6} T^3  \nonumber \\ 
 \pp _3 (T)&=&d\theta_0 + J_0(v_0\frac{1}{2} T^2 + \theta_0 \frac{1}{3} T^3 +  J_0 \frac{1}{8} T^4)
\label{p3}\\
\ff &=& \ff (\tau)\nonumber \\
\ff _+&=&\ff (\tau+T)  \nonumber  %
\end{eqnarray}
For a given function $\phi_1$, we  solved the system eq.~(\ref{eqtnT1}) for $\ff (\tau)$  and $T(\tau)$ as a delay-differential equation, using the matlab routine {\it ddesd}, after some transformation to convert it into standard form for which the delays are backwards. This amounts to solving for incoming electric field $\psi$ for given values of the potential difference $-\phi_1$.

We present numerical computation for a solution that starts in the steady state    $\bar F$ with  $(\bar v_0, \bar J_0, \bar \phi_1)=(0.5,1,-1)$ on a system with thickness $d=4/3$ (and with $\bar \psi_0=-0.5$), for $t<4$. This steady state is critical in that the potential difference is at its minimum (i.e., $-\phi_1=-\phi_{min}$) and the flux is at its maximum (i.e., $J_0=J_{max}$). The potential $\phi_1$ varied linearly over the time interval $4<t<5$ up to the value $\tilde \phi_1=\bar \phi_1 +0.2$, and then held constant at this value. Since this decreases the value of the potential jump $-\phi_1$, it does not lead to a  steady state solution.  

\begin{figure}
\begin{picture}(0.0,150.0)
\includegraphics{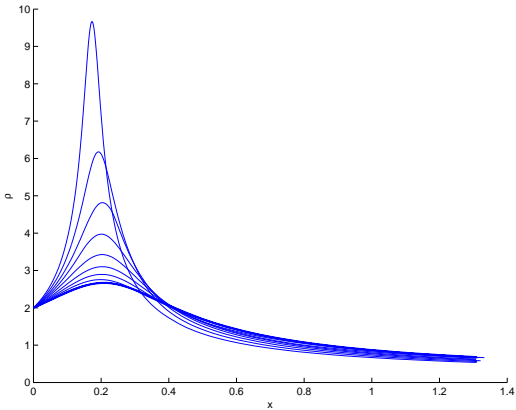}
\end{picture}
\caption{The density $\rho$ for steady boundary data for which  $J_0>J_{max}$. Values of $\rho$ are presented at 21 times starting at $t=0$ and at intervals of $0.4008$. Near $x=0.2$ value of $\rho$ is increasing monotonically as a function of $t$.}
\label{fig:RhoCase1Run2}
\end{figure}

\begin{figure}
\begin{picture}(0.0,150.0)
\includegraphics{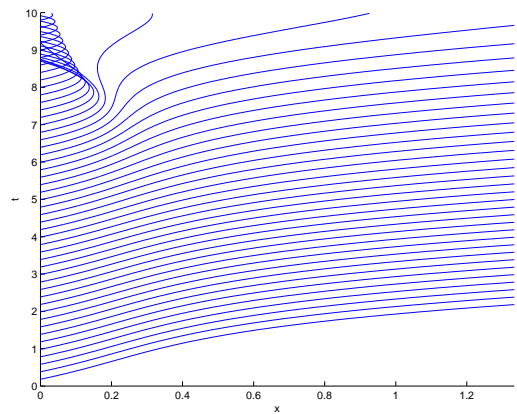}
\end{picture}
\caption{Characteristics (i.e., particle paths) for steady boundary data for which  $J_0>J_{max}$. The solution breaks down when there is a cusp in the characteristics.}
\label{fig:CharsCase1Run2}
\end{figure}

The resulting density $\rho$ is presented in Figure \ref{fig:RhoCase1Run2}, which shows development of a singularity. Nevertheless, the function $f'''$ remains smooth and bounded, so that implicit solution formulation remains valid up to the time of singularity formation.  Characteristics are shown in Figure \ref{fig:CharsCase1Run2}, which shows formation of a caustic.   Note that the velocity becomes negative before the singularity.

\begin{figure}[h]
\begin{picture}(0.0,150.0)
\includegraphics{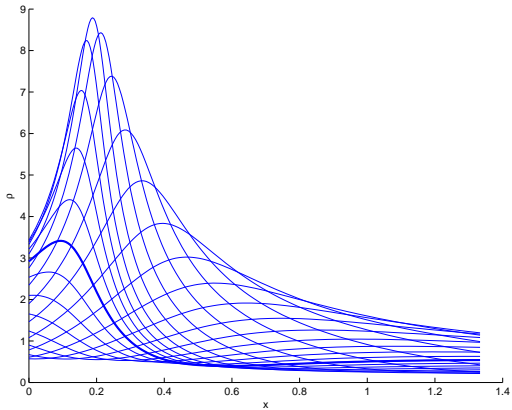}
\end{picture}
\caption{The density $\rho$ at various times for periodic boundary data for which  $\bar J_0> \bar J_{max}$. Values of $\rho$ are presented at 21 times starting at $t=0$ (bold curve) and at intervals of $dt=0.1612355$. The highest value of $\rho$ in this figure occurs at $t=5dt$,   the time of  bunching of characteristics in Figure \ref{fig:CharsPCase3Run2}. }
\label{fig:RhoPCase3Run2}
\end{figure}

\begin{figure}[h]
\begin{picture}(0.0,150.0)
\includegraphics{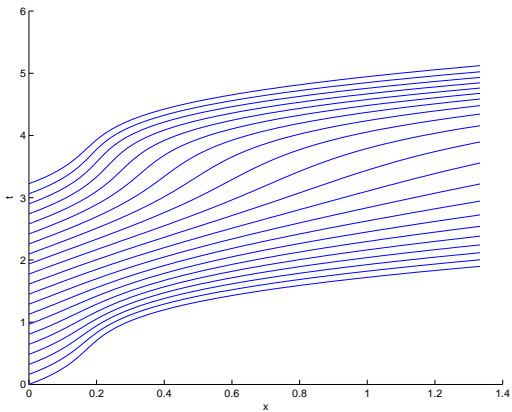}
\end{picture}
\caption{Characteristics (i.e., particle paths) for periodic boundary data for which  $\bar J_0> \bar J_{max}$. Note that a cusp nearly forms in the characteristics.}
\label{fig:CharsPCase3Run2}
\end{figure}

As a second example,  consider unsteady solutions  having constant incoming velocity $v_0$ but periodic flux $J_0(\tau)$ and periodic incoming electric field $\psi_0(\tau)$. The implicit form of the solution eq.~(\ref{charform})
 is given by 
\begin{eqnarray}
\psi(s,\tau)&=& \theta(\tau)+a_0 s +\ff '''(s+\tau)   \nonumber \\ 
 v(s,\tau) &=& v_0+\theta(\tau)s + a_0 \half s^2+\ff ''(s+\tau)-\ff ''(\tau) \nonumber \\ 
 x(s,\tau)&=& v_0 s+\half \theta(\tau)s^2+ a_0 \frac{1}{6} s^3 \nonumber \\
&& +\ff '(s+\tau)-\ff '(\tau)-\ff ''(\tau)s  \nonumber  
\end{eqnarray}
in which  $a_0$ is a constant, and $\theta$ and $\ff$ are prescribed periodic functions; i.e.,
\begin{eqnarray}
\ff(\tau)&=& \ff_1 \sin(k\tau/t_0)  \label{fperiodic}  \nonumber \\ 
\theta(\tau)&=&\theta_0+  \theta_1 \sin(k\tau/t_0+\tau_1)   \nonumber  
\end{eqnarray}
with period $P=2\pi t_0/k$. The incoming flux is
\begin{eqnarray}
J_0(\tau)  =a_0-\theta'(\tau) .  \nonumber  %
\end{eqnarray}

For given values of the constants $v_0$, $\theta_0$, $a_0$, $\ff_1$, $k$, $\theta_1$ and $\tau_1$, the solution is constructed numerically:
First, the crossing time $T(\tau)$ is found   by solving 
\begin{eqnarray}
x(T(\tau),\tau) =d , \nonumber  %
\end{eqnarray} 
and the potential $\phi_1=\phi(T(\tau),\tau)$ is found by  numerical computation of the integral  eq.~(\ref{phiIntegral}); i.e.,
\begin{eqnarray}
 \phi_1(\tau) =- \int_0^{T(\tau)} (v-\partial_\tau x) \psi (s',\tau+T(\tau)-s') ds'  \nonumber  %
\end{eqnarray}
for a discrete set of values of $\tau$.
Second, the average of the incoming current ${\bar J_0}$ and the  adiabatic  average of the maximal current ${\bar J_{max}}$ are numerically calculated as
\begin{eqnarray}
{\bar J_0} &=& P^{-1} \int_0^P J_0(\tau)d\tau =a_0   \nonumber \\ 
{\bar J_{max}} &=& P^{-1} \int_0^P J_{max}(\tau) (1+T'(\tau)) d\tau  \nonumber  
\end{eqnarray}
in which $J_{\max}(\tau)$ is defined by eq.~(\ref{Jmax}) using $\phi_1=\phi_1(\tau)$. 
Note that ${\bar J_0}$ is averaged over $s=0$ (i.e., $x=0$) where $dt=d\tau$ and  ${\bar J_{max}}$ is averaged over
$s=T(\tau)$ (i.e., $x=d$) where $dt=(1+T'(\tau))d\tau$. 
Finally, form the difference 
\begin{eqnarray}
J_{diff} &=& {\bar J_0} - {\bar J_{max}}.   \nonumber 
\end{eqnarray}

Solutions with $J_{diff}>0$ (i.e., that exceed the Child-Langmuir limit on average) were found by Monte Carlo search over the values of the parameters $k$, $\theta_1$, $\tau_1$. Results are shown below for $( v_0, a_0,\theta_0, \ff_1,k,\theta_1,\tau_1)=(0.5,1,-0.5,0.1,3.8969,0.36762,2.2684)$ on a system with thickness $d=4/3$.   In unscaled variables, the ratio of potential energy to the energy of the entering particles is approximately $6.5$ corresponding, for example, to $100 eV$ electrons on a $0.65 kV$ potential.
The density $\rho$ and characteristics are presented in Figures \ref{fig:RhoPCase3Run2} and \ref{fig:CharsPCase3Run2}.
The resulting average values are $\bar J_0=1 $ and $\bar J_{max} =0.85026$, so that the  average incoming current $\bar J_0$ exceeds the adiabatic average of the maximal current $\bar J_{max}$ by an amount $J_{diff}=0.14974$, an increase of about $17\%$.

Recent related work includes  evidence that the  average flux cannot exceed  $J_{max}$, under the additional constraints that $v_0=0$ and that $\phi_1$ is constant \cite{Griswold2010}, and   experimental and numerical results for short current pulses  that exceed $J_{max}$, as well as for increases in current beyond the formation of a virtual cathode  \cite{Lau2002}.

The unsteady solutions constructed above suggest that the implicit solution formulation may be  useful for exploring additional properties of the diode equations, such as solutions that maximize the electric field strength and control methods to prevent formation of virtual cathodes.

Thanks to J. Luginsland for helpful suggestions. Research supported in part by the Air Force Office of Scientific Research STTR program through  grant FA9550-09-C-0115 and by the Department of Energy through grant DE-FG02-05ER25710.

\end{document}